# The Exchange Bias effect in pure $Co_2C$ nanoparticles

Nirmal Roy[1], Md. Arif Ali[1], Arpita Sen[2], Prasenjit Sen[2,*], S. S. Banerjee[1,+]

[1]*Indian Institute of Technology Kanpur, Kanpur 208016, Uttar Pradesh India*

[2] *Harish-Chandra Research Institute, Homi Bhabha National Institute, Chhatnag Road, Jhunsi, Allahabad, Uttar Pradesh 211019, India*

**ABSTRACT:** We study the low temperature magnetic properties of nanoparticles of pure transition metal carbide, viz., $Co_2C$, with an average particle diameter of 40 ± 10 nm. These $Co_2C$ nanoparticles are ferromagnetic (FM) up to room temperature with blocking temperatures above room temperature. The coercive field shows abrupt deviation from the Kneller's law below 50 K. In this low temperature regime the magnetization hysteresis loop shows shifts due to exchange bias (EB) effect, with an exchange field of ~ 250 Oe. Analysis of training of the EB effect and ac and dc magnetic measurements suggest that EB arises in the nanoparticles due to a core-shell structure with a FM core and a cluster glass shell. The shell contains uncompensated spins, some of which are freely rotatable while some are frozen. DFT calculations of structural and magnetic properties of small $Co_2C$ clusters of diameter of few Angstroms confirm a core-shell structure, where the structurally ordered core has uniform magnetic moment distribution and the structurally disordered shell has non-uniform moment distribution.

**INTRODUCTION:** Interfacial exchange coupling between ferromagnetic (FM) and antiferromagnetic (AFM) materials leads to unidirectional exchange anisotropy, which pins the FM layer. [1,2] When such materials are cooled through the Neel temperature of the AFM component in presence of a magnetic field, a shift in the hysteresis loop is observed due to the exchange anisotropy across the interface pinning the FM layer. This is called exchange bias (EB) effect, first observed in Co/CoO by Meiklejohn and Bean. [1,3,4,5] The EB effect has been studied extensively due to its application to ultrahigh density memory devices, spin valves and giant magneto-resistance effect. [6,7,8,9,10,11,12,13] Apart from conventional FM/AFM interface, Ferrimagnet (FI)/FM, FI/AFM, FM/spin glass (SG) and some of the phase separated bulk materials of cobaltites and manganites also show EB effect. [14,15,16] At the nano-scale, there are examples like $NiFe_2O_4$ and NiO nanoparticles, which have a core with ordered moments, and a shell with disordered spin configuration. [17,18] FM/SG [16] core-shell structures with a FM core and a SG shell with disordered spin configuration exhibit EB effect. [19] Here the EB effect strongly depends on the cooling field (CF), as the SG state has multiple equivalent spin configurations associated with a multi-local minima's in the free energy landscape. [18,19] Observations suggest that depending on CF value which is applied above the glass transition temperature ($T_g$), the ordered phase magnetization gets aligned in the applied CF direction and is pinned by a shell of disordered spin configuration. [13,14,20] It is believed that the spin configuration in the SG phase below $T_g$ is exchange coupled through interfacial exchange interaction with the ordered component, leading to EB effect in these SG systems. [14]

Transition metal carbide systems (TMCs) have received attention in recent times due to their exotic magnetic and electronic properties, which have wide potential for application in memory devices, ferro-fluids, bio-imaging etc. [21,22,23] Amongst these TMCs, cobalt carbide is a potential candidate for rare earth free permanent magnets due to its high coercivity and remanence at room temperature, along with its cost-effective synthesis routes. [24,25] These TMCs can also be synthesized as nano-composites, giving them wider potential for applications. However being nano-sized, their ordering as well as blocking temperatures are often below room temperature, thereby restricting their application potential. Recently, investigations into the magnetic properties of $Co_3C$-$Co_2C$ nanocomposite showed possible high blocking temperature suggesting that TMC nanocomposite could be potentially useful.[26] These $Co_3C$-$Co_2C$ nanocomposite also exhibit unusual spin wave localization phenomena which gave rise of unusual temperature dependence of the coercive field.[26] These results suggeted a complex interaction between two components in the admixed nanocomposite state, viz., the $Co_3C$ phase which is a known ferromagnetic phase interacting with what seemed like a weakly magnetic $Co_2C$ phase, in the nanocomposite. This finding was a surprise as studies suggest bulk form of $Co_2C$ is paramagnetic.[27,28] It maybe therefore be worthwhile investigating the magnetic properties of $Co_2C$ nanoparticles. The pre-existing literature in this respect is limited in scope. There are reports on the behavior of coercivity of $Co_2C$ nanoparticles which is seen to vary from 450 to 1200 Oe as the temperature (*T*) is lowered, and low saturation magnetization of ~13 emu/g.[29] Beyond this there is not much exploration into unravelling the magnetism of $Co_2C$ nanoparticles. Here we explore in detail the low *T* magnetic properties of the pure $Co_2C$ nanoparticles. *M-H* hysteresis measurements show that $Co_2C$ is ferromagnetic up to room temperature with high blocking temperature. The behavior of the temperature dependence of the coercive field can be delineated into two separate regimes (I and II). Above 50 K (regime II) the temperature dependence of coercive field obeys the conventional Kneller's law. However, below 50 K (regime I), there is an abrupt deviation from Kneller's law. This study shows that the blocking temperature of the $Co_2C$ nanoparticles is atleast above room temperature. A careful study of *M-H* hysteresis loops at low *T* reveals horizontal shift in the loop which suggests the presence of EB-effect. Observation of magnetic training effect (TE) at 2 K further confirms the presence of EB-effect in these pure $Co_2C$ nanoparticles and the presence of frozen and rotatable spin glass state in the system. Detailed analysis of the temperature dependent features in ac susceptibility at different frequencies ($f$) shows the existence of a spin freezing temperature ($T_f$), and a low $T$ cluster glass state coexisting with the FM fraction in the $Co_2C$ nanoparticles. Our studies suggest that EB-effect in the $Co_2C$ nanoparticles is related to a FM and cluster glass core shell structure. We perform DFT calculations on small $Co_2C$ particles with diameters of a few Angstroms. These calculations show that the particles possess a structurally ordered core with atomic co-ordination similar to that in bulk $Co_2C$. The core is surrounded by a structurally disordered shell with under-coordinated atoms compared to the core. While the core has uniform magnetic moment distribution, the shell has non-uniform magnetic moment distribution. These calculations confirm the presence of a core-shell structure in $Co_2C$ at low dimensions.

## RESULT AND DISCUSSION

The X-ray diffraction (XRD) of powder of $Co_2C$ nanoparticles is show in figure 1a (see method for details). A comparison of the refined lattice parameters with the standard values (see Table 1), shows small deviation of the refined values from the standard ones. Figure 1b shows Scanning Electron Microscopy (SEM) image of the nanoparticles. Estimates based on the X-ray peak width and Transmission Electron Microscopy (TEM) study (see supplementary information, SI 1)) gives an average $Co_2C$ nanoparticle diameter of 40±10 nm.

**Table 1. Comparison of Reitveld refined lattice parameters of $Co_2C$ nanoparticles with standard values.**

| Lattice Parameter | Standard Value (Å) | | | Refined Value (Å) | | |
|---|---|---|---|---|---|---|
| | a | b | c | a | b | c |
| $Co_2C$ | 4.3707 | 4.4465 | 2.8969 | 4.3797 | 4.4632 | 2.9030 |

Figure 1c shows the ferromagnetic nature of the *M-H* hysteresis loop measured at different *T*. Here we would like to mention that whereas we observe the $Co_2C$ nanoparticles to be FM in nature, bulk $Co_2C$ is expected to be paramagnetic. [27,28] It is known that magnetic nanoparticles possess a characteristic blocking temperature $T_B = K_{eff}V / 25k_B$, where $K_{eff}$ is the effective anisotropy energy density and $V$ is the volume of a single magnetic domain, and $k_B$ is the Boltzmann constant. At $T > T_B$ the magnetic anisotropy energy barriers are overcome by thermal fluctuations and hysteresis in *M-H* is lost. The observation of a significant hysteresis (Figure 1c) persisting up to room temperature suggests that in $Co_2C$ nano-particles (~40 nm diameter) $T_B$ is larger than room temperature. Although $Co_2C$ nanoparticle has a substantial saturation magnetization ($M_S$) at room temperature of 6.1 emu/g, it is smaller than the value for pure Co and $Co_2C$-$Co_3C$ nanocomposite. [26] The behavior of the coercive field ($H_c$) (determined from *M-H* loops) as a function of *T* is shown in Figure 1d. The high blocking temperature seems to suggest presence of strong magnetic anisotropy in nanoparticles of $Co_2C$.

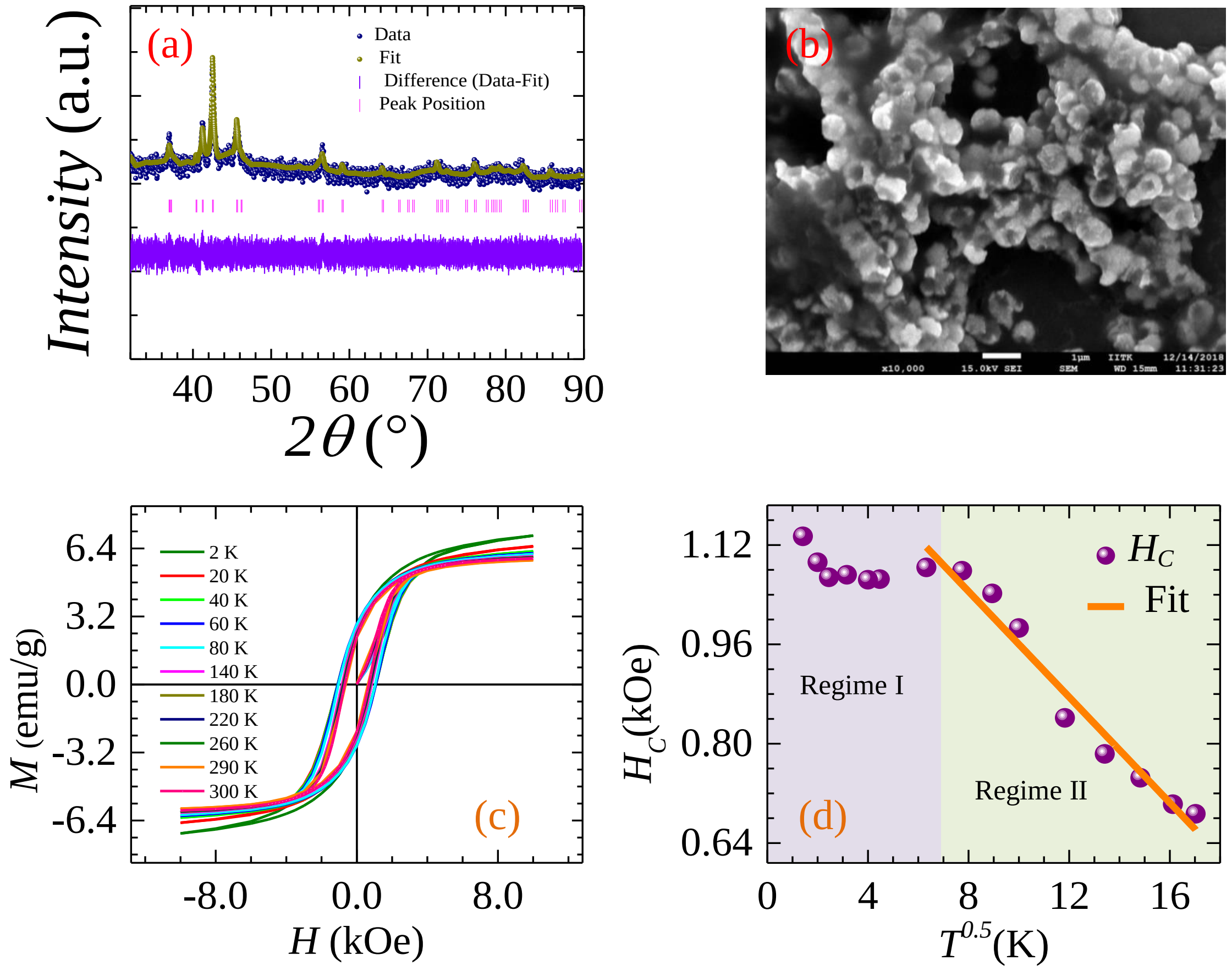

**Figure 1. (a) Powder X-Ray diffraction pattern (navy circle) of $Co_2C$ nanoparticles. The dark yellow line, short vertical bar and violet line represent the Rietveld refinement profiles, the fitted Bragg positions, and the residuals respectively. (b) SEM image nanoparticles. (c) Hysteresis loops of the pure $Co_2C$ nano-particles at different temperatures. (d) Plot of Coercive field ($H_C$) vs. $T^{0.5}$. Orange line is the straight-line fitting according to Kneller's law. Regime II follows the typical Kneller's law while in the low $T$ regime there is a significant deviation from the law (see text for details).**

For magnetic nanoparticles with uniaxial magnetic anisotropy, thermal activation of magnetic domains over the magnetic anisotropy energy barrier gives rise to the temperature dependence of $H_C$. The typical behavior of $H_C(T)$ is given by the Kneller's law, viz., $H_C(T) = H_0\left[1-\left(\frac{T}{T_B}\right)^{0.5}\right]$, where $H_0$=$H_C$($T$=0 K), and $T_B$ is the blocking temperature. In the regime II of Figure 1d, the plot of $H_c$ vs. $T^{0.5}$ shows good agreement with Kneller's law down to ~50 K (the boundary between regime I and II). Below 50 K, viz., in regime I, there is a significant deviation from Kneller's law. Slope and intercept of the straight line in regime II of Figure 1d give an estimate of $T_B$=1085±25 K, which is well above room temperature. We should mention that this is an overestimation of $T_B$ as Kneller's law for our system is valid only in a limited $T$ regime, and sometimes in literature exponent value different from 0.5 has also been used in the Kneller's law. [30] Despite our overestimating $T_B$, we know from our *M-H* hysteresis measurement (Figure 1c) that its lower bound for $T_B$ has to be greater than room $T$ as we have already noted a significant *M-H* hysteresis at room temperature. It is interesting to note that typical $T_B$ for 5 nm diameter $NiFe_2O_4$ [31] and NiO [32] nano-particles are 61 K and 5 K, respectively which rules them out for applications, and make $Co_2C$ nano-particles even more promising.

As the $T$ is reduced below 50 K (regime I of Figure 1d) the *M-H* hysteresis loop shows an asymmetric shift, both in $H$ (about zero) and in $M$ value. At higher $T$, viz., regime II, the shift is absent and the hysteresis loop is symmetric. Figure 2a shows the asymmetric nature of *M-H* loops at 2 K. The shift in *M-H* loops suggests the EB-effect. The EB field ($H_{EB}$), the EB magnetization ($M_{EB}$) and coercive field ($H_C$) are defined as, $H_{EB}=\left(\frac{|H_L|-|H_R|}{2}\right)$, $M_{EB}=\left(\frac{|M_1|-|M_2|}{2}\right)$, and $H_C=\left(\frac{|H_L|+|H_R|}{2}\right)$ where $H_L$ and $H_R$ are left and right coercive fields (see Figure 2a), and $M_1$ and $M_2$ are positive and negative magnetization values at $H$=0 Oe. In Figure 2b shows that $H_{EB}$ ~250 Oe at low $T$ (in regime I of Figure 1d), and it decreases exponentially to zero at 50 K (in regime II of Figure 1d). It may be noted that an exponential decrease of $H_{EB}(T)$ is characteristic of a FM/SG system. [33]

An important test of EB effect is to show the presence of magnetic training effect (TE). In TE, a system exhibiting EB effect is first field cooled in a biasing field (training field). Then, a certain number of complete cycles of the $H$ field decreases its anisotropy significantly, leading to a decrease in $H_{EB}$. For the $Co_2C$ nanoparticles, we have used a training field of 1 T applied at 50 K (viz., the $T$ where $H_{EB}$ = 0). We have measured thirteen consecutive $M$-$H$ loops at 2 K. We see that after a certain number of cycles of $H$, $H_{EB}$ converges to a fixed value of ~ 0.12 kOe (see Figure 2c and 2d). The TE is a macroscopic fingerprint of the configurational rearrangement of the uncompensated AFM spin towards equilibrium in FM/AFM system. [19] The AFM state could also be a SG like state, in which there exist multiple equivalent spin configurations due to its multi-valley energy landscape. [19] The TE could be related to configurational rearrangements of spins in the SG like state.

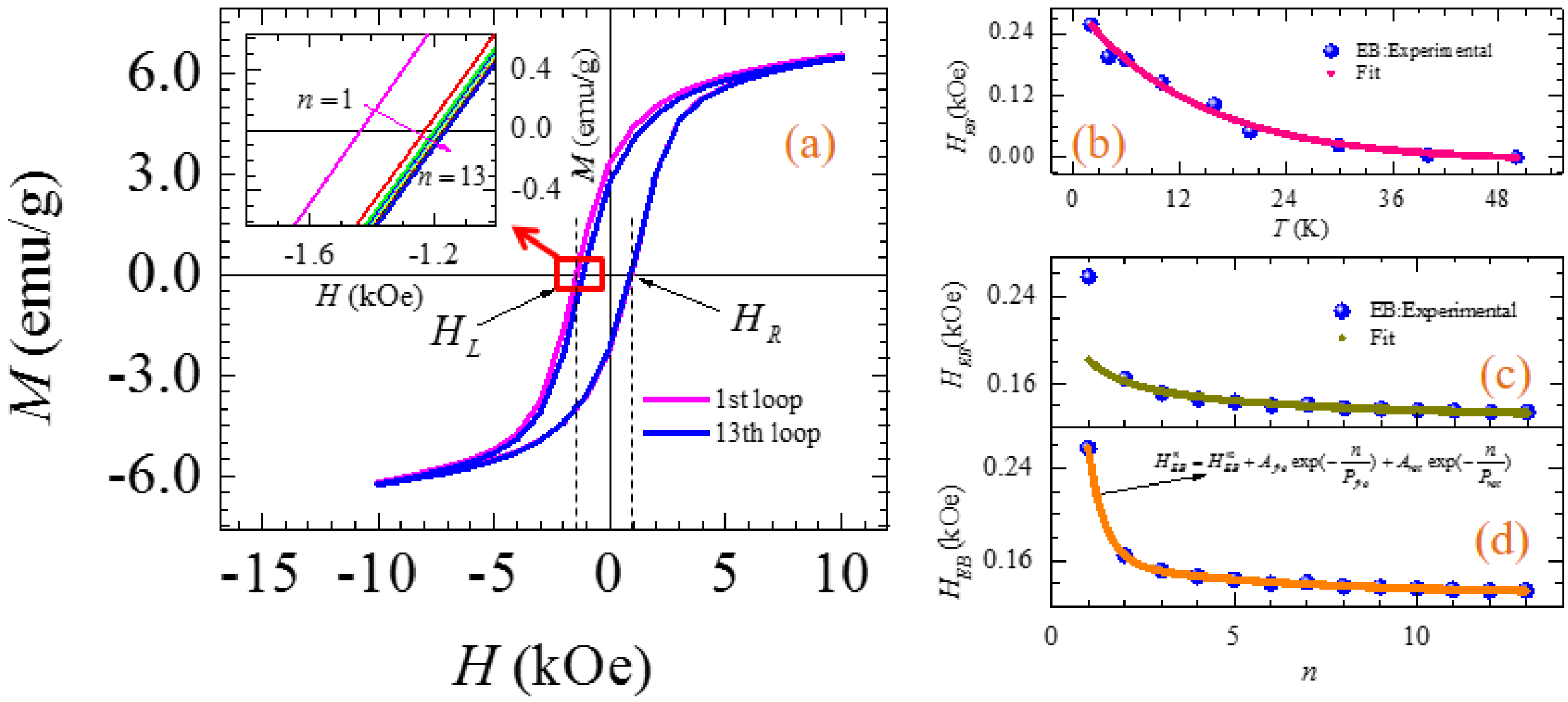

**Figure 2. (a) *M-H* loop for 1st and 13th training cycles at 2 K with FC of 1T. Inset of the Figure 2a shows the expanded view of a leg of the *M-H* loop showing magnetic training effect at 2 K. With number of loops *n*, a significant shift is observed in $H_L$ (see inset) while the shift in $H_R$ is comparatively smaller (main panel). This clearly suggests a change in $H_{EB}$ with *n* and the presence of TE in $Co_2C$ nanoparticles. (b) Temperature dependence of $H_{EB}$. The data is fitted to exponential fit of the form $H_{EB} = H_0 \exp(-\frac{T}{T_1})$ with $H_0$=291±14 Oe, and $T_1$=13±1 K. $H_{EB}$ is calculated as the average of $H_{EB}$ obtained with positive and negative training field. (c) $H_{EB}$ as a function of the loop index (*n*). Solid circle represents an experimental da-**

**ta, and solid line is fitted with equation $H_{EB}-H_{\infty}=\frac{k_H}{\sqrt{n}}$. (d) $H_{EB}$ vs. $n$ plot. Solid line is the fitting with frozen and rotatable spin relaxation model. The data is fitted to a form shown in Figure (see text for details).**

Figure 2a shows the 1st and the 13th complete loops (±1T). We plot $H_{EB}$ as a function of loop index ($n$) in Figure 2c and Figure 2d. Behavior of $M_{EB}$ vs. $n$ is shown in supplementary information section, SI 2. It is clear that $H_{EB}$ (and $M_{EB}$) decreases with $n$ showing TE, thereby confirming the presence of EB effect. [34,35] Figure 2c shows that $H_{EB}(n)$ fits to a well-known form [36] $H_{EB}-H_{\infty}=\frac{k_H}{\sqrt{n}}$ only for $n\geq 2$, where $H_{\infty}$ is the EB field in the limit of $n\rightarrow\infty$, and $k_H$ is a system-dependent constant which is proportional to the exchange bias anisotropy energy scale. SI 2 shows that $M_{EB}$ follows the behavior $M_{EB}-M_{\infty}=\frac{k_M}{\sqrt{n}}$ for $n\geq 2$. The fits give $H_{\infty}=114\pm 1.5$ Oe, $M_{\infty}=2.71\pm 0.01$ emu/g, $k_H$ = 67±3 Oe and $k_M$ = 0.25±0.02 emu/g. We note that the values of $k_H$ and $k_M$ are comparable to that reported for EB effect found in FM / SG system. [19] Note that the above form does not fit the steep decrease in $H_{EB}$ from its value at $n=1$. In supplementary section SI 3, we show that an iterative model obeyed by EB effect, viz., Binek model, [37] also fits our $H_{EB}(n)$ behavior only for $n\geq 2$.

Recently, an alternate explantation of TE has been given for EB seen in FM/SG systems. In this model, the SG state is considered to have both frozen and freely rotatable uncompensated spins. [38] The FM component is exchange coupled to both these components in the SG layer. The two components possess very different relaxation rates and contribute to different extents to the TE of $H_{EB}$. The following equation has been proposed for the $n$ dependence of $H_{EB}$.

$$H_{EB}^{n}=H_{EB}^{\infty}+A_{fro}\exp(-\frac{n}{P_{fro}})+A_{rot}\exp(-\frac{n}{P_{rot}}), \quad (1)$$

where $A_{fro}$, and $P_{fro}$ are parameters related to the frozen component, and $A_{rot}$ and $P_{rot}$ are parameters corresponding to the rotatable component of the spin at the interface. The parameters *A's* have dimensions of magnetic field (Oe) while *P's* are dimensionless numbers, but are taken to be analogous to relaxation time scale, and the time variable is replaced by the discrete training sequence number *n* in equation (1). Unlike the other models, the solid line in Figure 2d shows an excellent fit to the experimental data with equation (1) for all *n* values. The best fit parameters are, $H_{EB}^{\infty} = 131 \pm 5\, Oe$, $A_{rot} = 879 \pm 55\, Oe$, $P_{rot} = 0.45 \pm 0.01$, $A_{fro} = 34 \pm 5\, Oe$, and $P_{fro} = 5 \pm 0.1$. Fitting parameters show predominant contribution at the initial stage of training from a fast relaxation of the rotatable component ($P_{rot}$) of the uncompensated spin at the interface. The frozen component relaxes slowly ($P_{fro}$), and governs the behavior for larger *n*. Values of $P_{rot}$ and $P_{fro}$ show that the relaxation of the frozen component is 10 times slower as compared to rotatable component. It is worthwhile mentioning here that a similar feature has been reported for other spin glass systems as well. [38,39,40] From these results we suggest that the $Co_2C$ nanoparticle has a core-shell like structure. The core shell structure consists of a FM core which gives rise to *M-H* hysteresis. This FM core is surrounded by a shell with disordered SG-like spin configuration. The disordered shell has both rotatable and frozen uncompensated spin components which are exchange coupled to the FM core giving rise to EB effect.

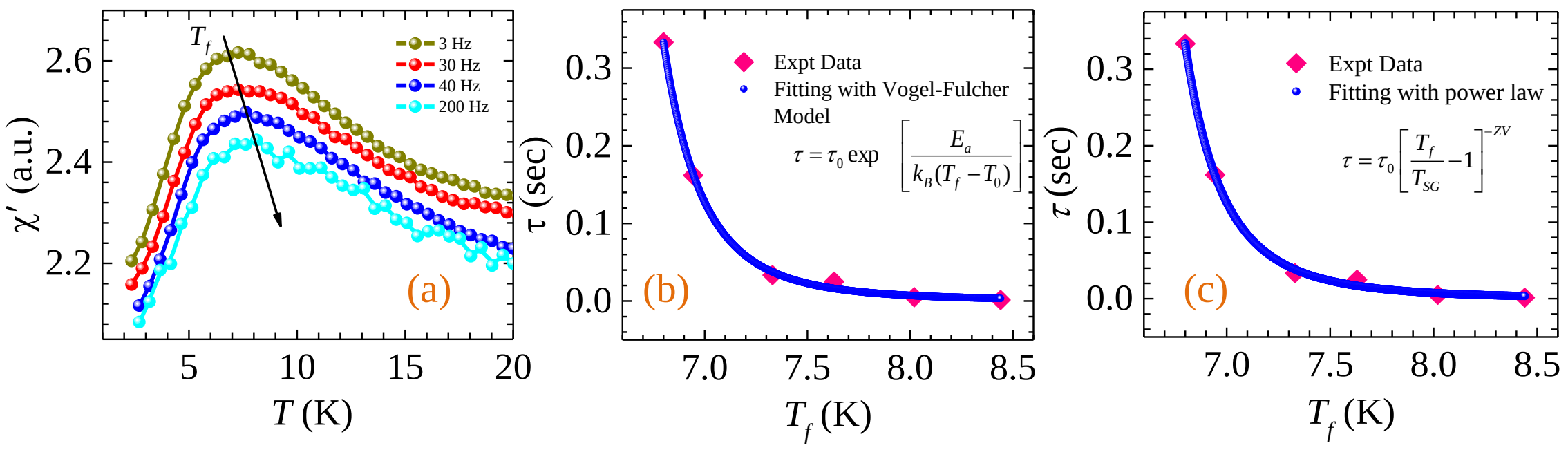

**Figure 3. (a) Real part of magnetic ac susceptibility with temperature. (b) Relaxation ($\tau$) as a function of freezing temperature ($T_f$). Solid line is the fitting with the Vogel- Fulcher model. (c) Relaxation ($\tau$) as a function of freezing temperature ($T_f$). Solid line is the fitting with dynamic scaling exponent associated with relaxation in SG system (see text for details).**

To confirm the presence of SG state, we measure the frequency-dependent in phase component of ac magnetic susceptibility, $\chi'(T)$. Figure 3a shows $\chi'(T)$ at different frequencies in zero dc applied field. The $\chi'(T)$ clearly shows a peak at a characteristic temperature $T_f$, which is a signature of spin glass freezing temperature. [41] Furthermore, as a SG system has multiple relaxation time scales, $T_f$ (maximum in $\chi'(T)$) shifts towards higher $T$ with increasing $f$. [41,42] For a SG system, the frequency dependent shift of the real part of ac susceptibility is expressed by the frequency sensitivity factor $K$, [43] defined as

$$K = \frac{\Delta T_f}{T_f \Delta\, log(\omega)} \quad . \tag{2}$$

$\Delta T_f$ and $\Delta(\log \omega)$ are the shifts of freezing temperature $T_f$ and $\omega = 2\pi f$. The typical range [44,45] of $K$ for SG systems is $10^{-2}$ to $10^{-3}$ and, $K \geq 0.2$ for the super-paramagnetic state. [45,46] The value of $K = 0.08$ obtained from equation (2) suggests that $Co_2C$ has a SG fraction. In Figure 3b and Figure 3c, using the data of Figure 3a, we plot the timescale corresponding to the frequency at which $\chi'(T)$ peaks, viz., $\tau = \frac{2\pi}{\omega}$ vs. $T_f$. At a first place, the $\tau$ vs. $T_f$ plot fits to the well-known Vogel-Fulcher law for SG systems, [47] $\tau = \tau_0 \exp\left(\frac{E_a}{k_B(T_f - T_0)}\right)$, where $\tau_0$, $E_a$ and $T_0$ are the characteristic magnetic relaxation time scale, activation energy, and Vogel-Fulcher temperature respectively. $T_0$ is a measure of the interfacial interaction energy in the SG system (interface between the FM and SG component present in the system). The best fit to the data in Figure 3b is obtained for $\tau_0 = (2.27 \pm .02) \times 10^{-5} s$, $E_a = 1.49 \pm 0.05\, meV$, and $T_0 = 5 \pm 0.06\, K$. In supplementary section SI 4 the analysis of time dependence of the relaxation of saturation magnetization also confirms the presence of a glassy state in $Co_2C$. Here, it must be mentioned that compared to conventional SG systems, where $\tau_0$ is in the range $10^{-12}\ to\ 10^{-15}$ s, [41] we find much slower relaxation rates. We believe the above difference shows that the origin of the SG state in $Co_2C$ is related to a disordered configuration of clusters where each cluster of atoms has a large net magnetic moment, viz., the so-called cluster spin glass. The relaxation of these disordered clusters is known to be slower compared to conventional SG system.[48] We perform a dynamical scaling analysis using critical slowing down model of the form [49,50]

$$\frac{\tau}{\tau_0} = \left[\frac{T_f - T_{SG}}{T_{SG}}\right]^{-z\gamma} \quad (3)$$

where $T_{SG}$ is the glass transition temperature, $z\gamma$ is the dynamic critical exponent, and $\tau_0$ is the spin relaxation time. An excellent fit to the experimental data with equation (3) is shown in Figure 3c and we confirm the presence of SG state in $Co_2C$ nanoparticles with a SG transition temperature ($T_{SG}$) of 6.15 ± 0.05 K. The fit gives $\tau_0 = (9.6 \pm 0.01) \times 10^{-5}$ s and $z\gamma = 3.6 \pm 0.1$, which is nearly the same as that reported for other cluster glass systems. [51] For conventional SG systems, the typical value [41] of $z\gamma$ ranges from 4 to 12 and $\tau_0 \sim 10^{-12}$-$10^{-15}$ s. The high value $\tau_0$ ($\sim 10^{-5} s$), we observe in $Co_2C$ has also been found in other systems, where cluster glass state has been reported. [44,51,52] The $z\gamma$ value of less than 4 supports our earlier result that the $Co_2C$ nanoparticle possesses a cluster glass state. Thus our results suggest that $Co_2C$ nanoparticle systems have a FM core with a cluster glass shell-like structure. Across the core shell interface the two magnetic phases are exchange coupled, leading to an EB effect.

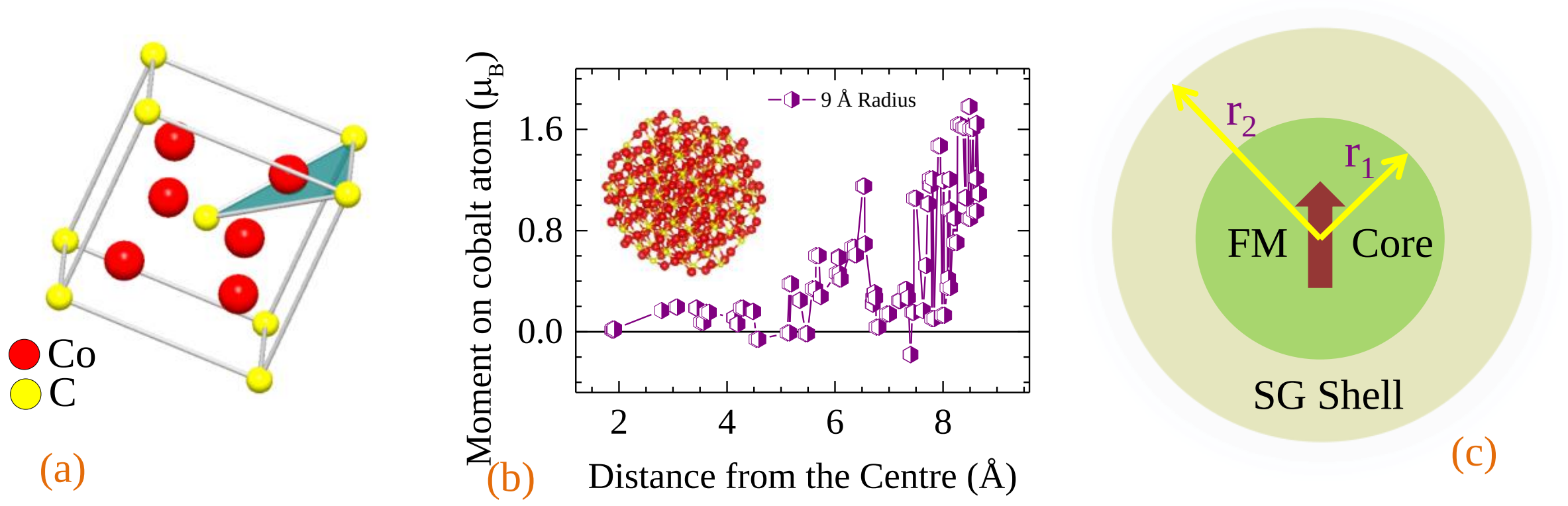

**Figure 4. (a) shows unit cell of $Co_2C$ with Co atom having 3 nearest neighbor C atom. (b) shows the theoretically calculated magnetic moment with distance measured from the center (of sphere located on a C atom) of the $Co_2C$ sphere of radius 9 Å. Inset shows the distribution of Co and C atoms of $Co_2C$ in the sphere of radius 9 Å. Color code for atoms is same as that in Figure 4a. (c) Schematic of core- shell structure of $Co_2C$ nanoparticles (See text for details).**

In order to further understand properties of $Co_2C$ nano-particles, we calculate their electronic structure using first-principles approach based on density functional theory (DFT). Studying 40 nm nanoparticles in DFT is computationally prohibitive. Therefore, we study properties of smaller particles that are computationally tractable. First, a $10\times10\times10$ (of the bulk unit cell) fragment of $Co_2C$ is constructed with the refined lattice parameters given in Table I. A sphere with the desired radius is then cut out from this, and its properties are calculated. For detailed methodology see supplementary section SI 5. In bulk $Co_2C$, each cobalt atom is surrounded by three nearest neighbor (NN) carbon atoms as shown in Figure 4a. In our calculations, we find a core-shell like structure. All Co atoms in the core are fully coordinated as in the bulk. However, in the outer shell the Co and C atoms no longer have bulk-like coordination.

We calculated properties of spherical particles of three different radii: 7 Å, 9 Å and 10 Å. Interestingly, all particles exhibit magnetic ground states with all the moment on the Co atoms. C atoms carry negligible moment. This should be contrasted with bulk $Co_2C$ which is paramagnetic. [27,28] The moment on the Co atoms is found to be a strong function of their distance from the center of the particle. It is about 0.014 $\mu_B$ at 1. 9 Å from the center, and then it increases, albeit with fluctuations. In order to check the radius dependence of the spin distribution, we studied a particle of 9 Å radius (see Figure 4b). We see that within about 4 Å from the center, all the Co moments are positive with an average value of ~ 0.1$\mu_B$. Beyond 4 Å the Co moments are quite large and also they fluctuate as one moves further outward. From this feature it seems that near the surface of the 9 Å sphere, the moment distribution is non-uniform and the value fluctuates as one moves deeper into the shell. The core, although on the average has a smaller moment, has less fluctuations in the moment distribution compared to the shell. Thus one has a core-shell like structure with a small but uniform and stable moment in the core and disordered moment configuration in the outer shell of the sphere. In a 7 Å and 10 Å sphere (see SI 6) we see similar features.

We now analyze the structure of the clusters of radii 7 Å, 9 Å and 10 Å. First, we identified the radial distance ($r_1$) from the central C atom up to which all the Co atoms have bulk-like $Co_2C$ coordination, viz., each Co atom has 3 NN C atoms. In the shell between $r_1$ and the boundary of the cluster ($r_2$) (see schematic in Figure 4c), we count the number of Co atoms having 1, 2 and 3 NN C atoms (denoted by $n_1$, $n_2$ and $n_3$) in the final relaxed structures. For the cluster with 9 Å radius, $n_1$, $n_2$ and $n_3$ are 23, 55 and 12. Thus the fractions of Co atoms having 1, 2 and 3 NN C atoms in the shell ($r_2$-$r_1$) are 0.26, 0.61 and 0.13 respectively (see supplementary section SI 7 for 7 Å and 10 Å sphere and more details on these parameters are shown in table 1 of SI 7). From this, it is clear that the shell between $r_1$ and $r_2$ has atoms with very different coordination compared to the core which has bulk coordination. The fluctuations in magnetic moment begin in the vicinity of the structurally disordered shell and extend into the shell as well. It seems that the changing spacing between atoms in the shell compared to the core has an effect on stabilization of the moments on the Co atoms. While more detailed investigations are needed to understand the above features, it appears that appropriate spacing of the Co atoms in $Co_2C$ structure, which governs the overlap of the atomic orbitals, is a crucial ingredient in stabilizing the moment on Co. We thus observe a core-shell like structure where there is a FM like core and a disordered spin configuration in the shell. We believe that in our $Co_2C$ nanoparticles, a core-shell structure is also present due to similar reasons as in the small sized particles. Exchange anisotropy present across such a core shell interface leads to the EB effect we observe. [17,19]

**CONCLUSION**

In conclusion we have studied magnetic properties of pure $Co_2C$ nanoparticles with typical diameter of 40 nm. These particles have large magnetic anisotropy which leads to hysteresis in magnetization surviving up to room temperature. An unusual temperature dependence of the coercive field of the system is found at low temperatures. At low temperatures we observe the emergence of exchange bias effect. Analysis of the ac and dc magnetic response of these nanoparticles shows that they possess a ferromagnetic component along with a spin glass component. The spin glass component is further shown to be like a cluster glass state with a fraction of spins which are freely rotatable and a fraction which are frozen. DFT studies of $Co_2C$ nanoparticles having diameters of a few Angstroms show a core-shell like structure. The core has almost uniform moment on the Co atoms with bulk-like coordination. However, in the shell, the co-ordination of the atoms is completely different. Magnetic mo-ments are also non-uniform in this structurally disordered shell. We believe the larger $Co_2C$ nanoparticles in our experiments possess similar core shell structures, where the structurally ordered core is a FM and a thin structurally disordered shell is a glassy state. The EB effect arises from such a FM-cluster glass core shell structure. The ferromagnetism observed in these TMC nanoparticles at room temperature along with their large blocking temperatures make them potentially attractive candidates for use as high density nano-magnetic memory and recording elements. They could also be useful for making non rare-earth based permanent magnets. Further studies are underway to unravel the rich and complex magnetic behavior of these nanoparticles.

## ASSOCIATED CONTENT

**Supporting information:** Additional supplementary information on TEM image of $Co_2C$, training of EB effect in magnetization, fitting training of EB effect to Binek's model, behavior of magnetization relaxation with different thermomagnetic history of the sample, details of DFT calculation details on small $Co_2C$ particles.

## METHOD

**Sample Synthesis:** The synthesis protocol of cobalt carbide nanoparticles is described in our previous work.[26] With respect to earlier synthesis route,[26] we have only varied the amount of NaOH to enable us to synthesize $Co_2C$ nanoparticles. Using 1.02g of NaOH, we successfully synthesized a pure phase of $Co_2C$ nanoparticles. The extracted pure $Co_2C$ nanoparticles were dried in vacuum and the powder compacted into pellets. Pellets of 6.2 mg were used for magnetic measurements.

**Physical Measurement:** Powder X-ray diffraction (XRD) was performed on the as synthesized powder of nanoparticles using Panalytical X-ray diffractometer with Cu $K_\alpha$ radiation (λ=1.5406 Å) at room temperature in the 2-theta range of (30°-90°). Reitveld refinement of the powder XRD pattern is carried out using Fullprof package indicated the presence of $Co_2C$ phase. SEM image of the nano-particles is captured with FESEM, JSM-7100F; JEOL. For TEM analysis, we first dissolved extremely little amount of powder of nanoparticles in to ethanol in micro centrifuge tube. Then After 10 mins of ultra-sonication of the solution, using micro-pipet, we put a small drop of suspension on the centre of carbon coated TEM grid and dry at room temperature for 30 mins. The prepared TEM grid is loaded in to FEI Titan G2 60 -300 for capturing TEM image. DC magnetization and ac – susceptibility measurements were performed on Cryogenics (UK) SQUID magnetometer. The background dc-magnetization response, of the sample holder was measured at the temperatures and frequencies at which the sample was measured. This background was subtracted from the sample magnetization response.

**Calculations:** All calculations were performed within the framework of plane-wave density functional theory (DFT). The Viennea *ab initio* simulation package (VASP) was used for the calculations. Energy cut off for the plane-wave basis set was 600 eV. Interaction between the valence electrons and the ion cores was represented by the projector augmented wave (PAW) potentials. The exchange-correlation energy was calculated employing the gradient-corrected functional proposed by Perdue, Burke and Ernzerhof (PBE). Brillouin zone integrations were performed using the **Γ-**point only.

## AUTHOR INFORMATION

**Corresponding Authors:**

[*]Email: prasenup@gmail.com

[+]Email: satyajit@iitk.ac.in

**Author Contributions:** All authors have equal contributions.

**ACKNOWLEDGEMENTS:**

SS Banerjee would like to acknowledge funding support from IIT Kanpur and Department of Science and Technology- India, AMT and Imprint II programs. The research of Arpita Sen was supported in part by the Infosys scholarship for senior students. All DFT calculations were performed at the HPC cluster computing facility at HRI (http://www.hri.res.in/cluster/).

**The Exchange Bias effect in pure $Co_2C$ nanoparticles**

Nirmal Roy[1], Md. Arif Ali[1], Arpita Sen[2], Prasenjit Sen[2,*], S. S. Banerjee[1,+]

[1]*Indian Institute of Technology Kanpur, Kanpur 208016, Uttar Pradesh India*

[2] *Harish-Chandra Research Institute, Homi Bhabha National Institute, Chhatnag Road, Jhunsi, Allahabad, Uttar Pradesh 211019, India*

---

Supplementary Information section SI 1:

TEM image of synthesised nanoparticle is shown below to determine the average particle size. The average particle size is ~40 nm.

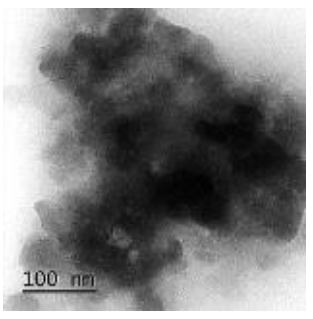

100 nm

**Figure: TEM image of the particles showing average size of the particles ~40 nm.**

Supplementary Information section SI 2:

Exchange Bias magnetization is defined as $M_{EB} = \frac{|M_1| - |M_2|}{2}$ where $M_1$ and $M_2$ are positive and negative remanent magnetization. The behaviour of $M_{EB}$ as a function of loop index ($n$) is shown in Figure below. $M_{EB}(n)$ follows the power law behavior of $M_{EB} - M_\infty = \frac{k_M}{\sqrt{n}}$ for $n \geq 2$, where $M_\infty$ is the EB magnetization in the limit of $n \to \infty$ and $k_M$ is a system dependent constant. From fit we get $M_\infty = 2.71 \pm .01$ emu/g and $k_M$=0.25±.02 emu/g.

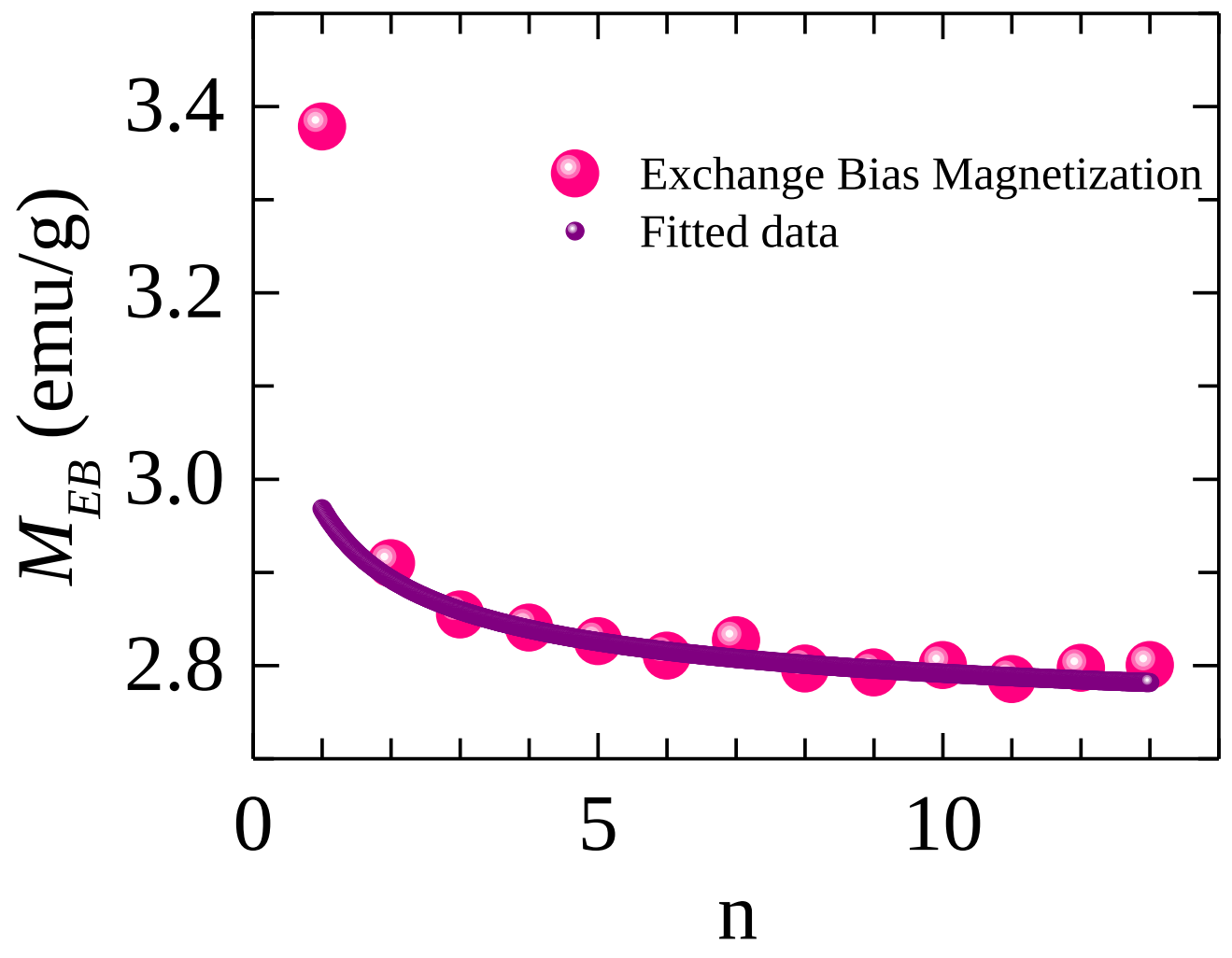

**Figure: $M_{EB}$ as a function of the loop index ($n$). Solid circle represents an experimental data, and solid line represents the best fit line to data using $M_{EB} - M_\infty = \frac{k_M}{\sqrt{n}}$.**

Supplementary Information section SI 3:

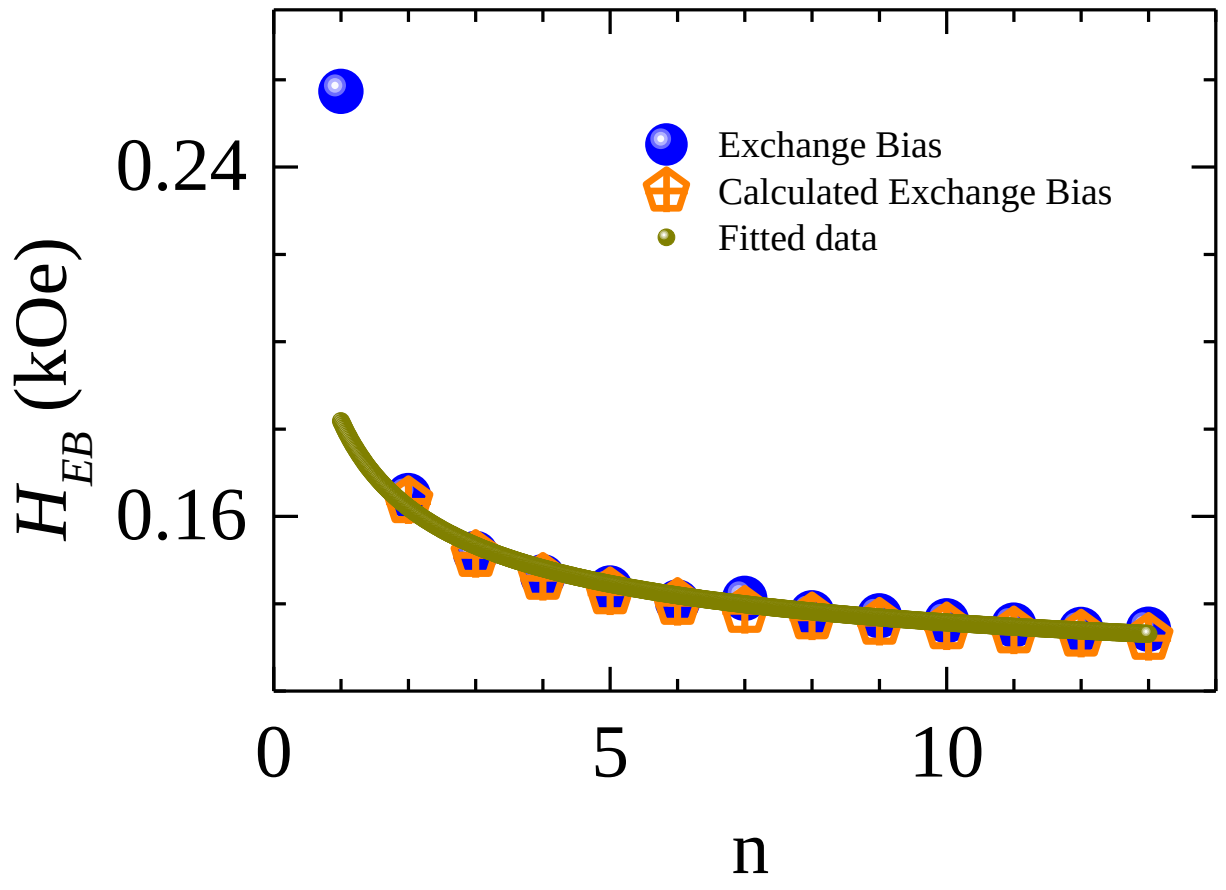

**Figure: $H_{EB}$ as a function of the loop index (*n*). Solid circle represents an experimental data, open cross circle represents $H_{EB}$ values calculated following the iterative Binek's model and the solid line is fitted with equation** $H_{EB} - H_{\infty} = \frac{k_H}{\sqrt{n}}$.

In Figure above, we use the iterative Binek model [1] of spin configurational relaxation which gives an iterative relation between $(n+1)^{th}$ loop $H_{EB}$ value to $n^{th}$ loop $H_{EB}$ value as

$$H_{EB}(n+1) = H_{EB}(n) - \gamma\left[H_{EB}(n) - H_{EB}(\infty)\right]^3 \qquad (1)$$

where $\gamma$ is constant which depends on sample, is related to spin damping constant of the system. During the consecutive field cycling process some of the frozen spin in SG state may flip along the direction of applied magnetic field and fall into other metastable configuration leading to decrease in $H_{EB}$. Orange cross open circle in Figure shows the theoretically calculated $H_{EB}$ for $n \geq 2$ using $\gamma = 1.2667 \times 10^{-5}$ and $H_{EB}(\infty) = 62.35$ Oe, matching well with the experimental data. The $\gamma$ value reported in literature [2] for EB found in spin glass system is ~$10^{-5}$ which is comparable to our data. Binek model explains our data well for $n \geq 2$ but it does not explain the *n*=1 data.

Supplementary Information section SI 4:

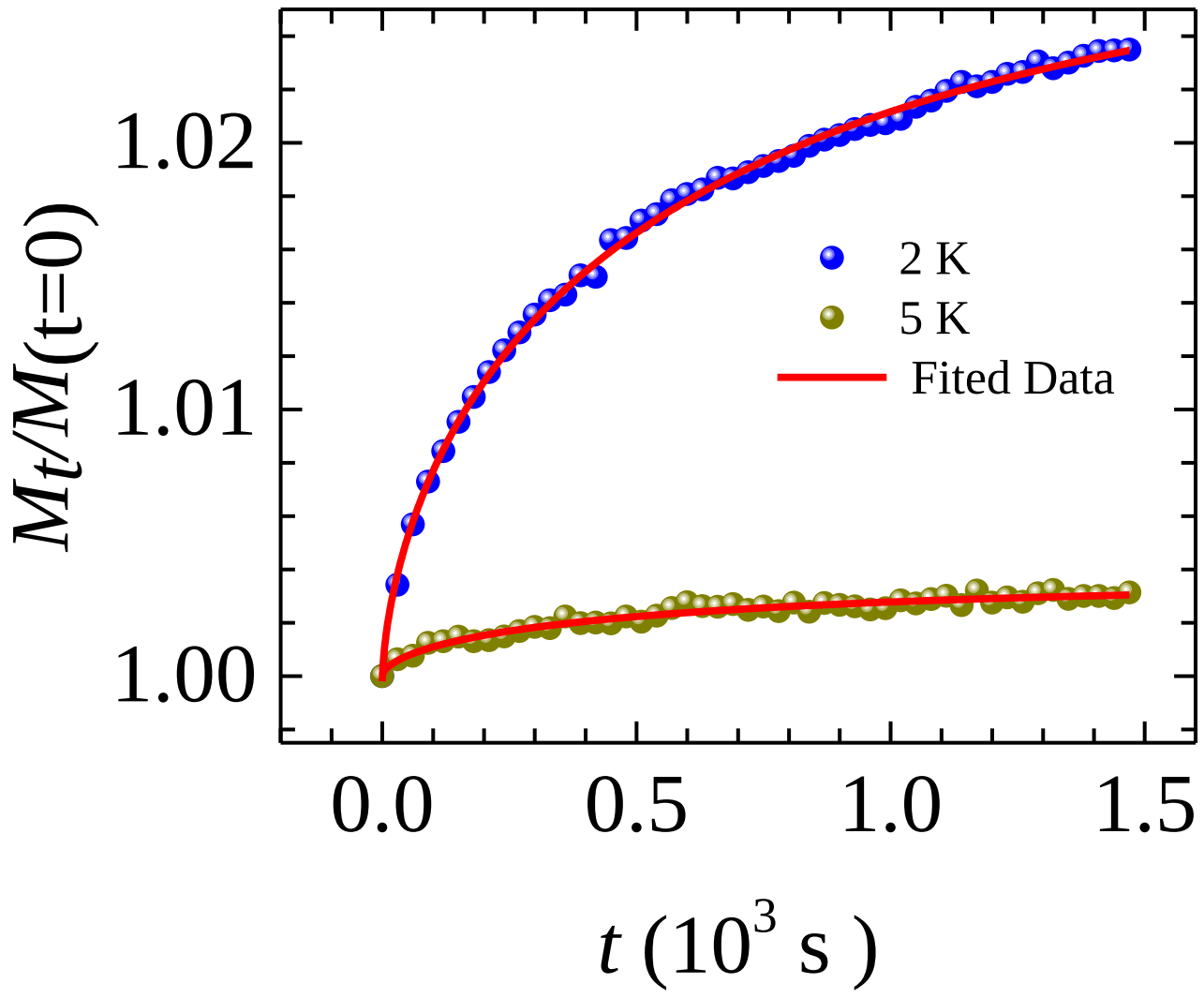

**Figure: Zero Field Cooled (ZFC) magnetization relaxation (normalized with respect to the magnetization at *t* = 0) measured at *T* = 2K (blue) and 5K (dark yellow) with *H* = 0.5T. The solid lines represent the fit with the stretched exponential expression given in text**.

To further study the dynamics of glassy state, we have measured the time dependent magnetization after ZFC from well above $T_f$ to measuring temperature with applied field 0.5T. Figure shows ZFC magnetization as a function of time at a temperature 5 K (near $T_{SG}$) and 2 K (far below the $T_{SG}$). Recall $T_{SG} \sim$ 6.15K. Magnetization doses not saturate even after a long time at 2 K because of randomly frozen of magnetic moment in the glassy state and it will take long time to aligned those frozen spin along field direction. We have fitted the time dependent magnetization data with standard stretched exponent function given below

$$M_t(H) = M_0(H) + \left[M_\infty(H) - M_0(H)\right]\left[1 - \exp\left\{-\left(t/\tau\right)^\alpha\right\}\right] \qquad (2)$$

where $M_0$ and $M_\infty$ are the magnetization at $t=0$ and $t \to \infty$ respectively, and $\tau$ is the characteristic relaxation time. We have obtained the best fit to the experimental data using $\frac{M_\infty}{M_0} = 1.029$ , $\tau = 657 \pm 17$ s, and $\alpha = 0.62$ for 2 K, and $\frac{M_\infty}{M_0} = 1.003$, $\tau = 507 \pm 66$ s, and $\alpha = 0.62$ for 5 K. The value of $\tau$ at 2 K is far larger than that of 5 K which is expected because 5 K is very close to $T_f$ and its value depends on how deep into the frozen state we are measuring. This once again confirmed the glassy state in $Co_2C$ nano-particles. Similar behaviour is also observed in other glassy system also.[3]

Supplementary Information section SI 5:

For the density functional theory (DFT) calculations, the exchange-correlation energy was calculated employing the gradient-corrected functional proposed by Perdue, Burke and Ernzerhof (PBE). [4] Brillouin zone integrations were performed using the **Γ-**point only. Spherical nano-particles of $Co_2C$ were formed using the following procedure. A 10×10×10 supercell of bulk (orthorhombic) $Co_2C$ was created, in which a C atom was found at the center. A sphere of the required radius, centered at the C atom, was cut out of this bulk fragment. The spherical nano-particle was then put inside a large cubic box, and its structure was relaxed. Vacuum along each of the three directions was kept 15 Å or higher to ensure that there are no interactions between the cluster and its periodic images. During these calculations, the energy and force convergence criteria were 0.0001 eV and 0.001 eV/Å respectively. The Viennea *ab initio* simulation package (VASP) [5,6,7,8] was used for all the calculations.

Supplementary Information section SI 6:

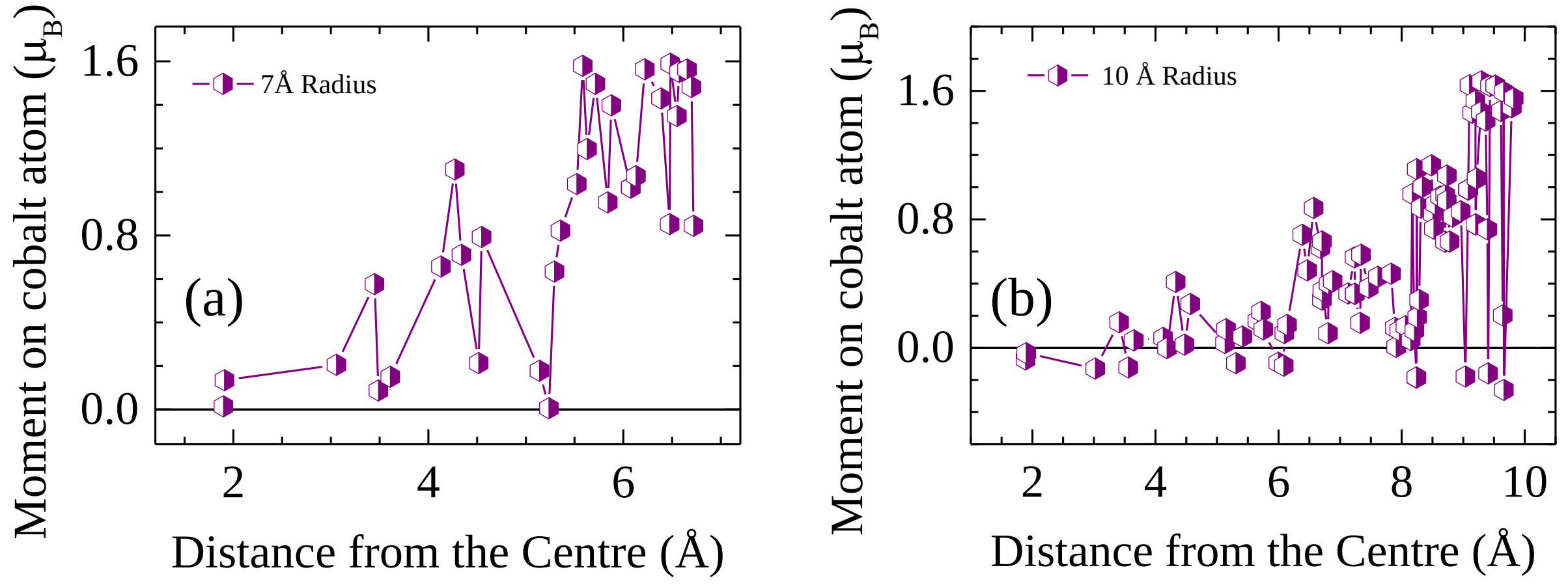

**Figure: Theoretically calculated magnetic moment on the cobalt atom in pure $Co_2C$ nano particles of radius (a) 7 Å and (b) 10 Å.**

Supplementary Information 7 (SI 7):

Table 1: 10 Å, 9 Å and 7 Å $Co_2C$ particles relaxed structure

**Radius 10 Å Relaxed Stucture**

| ($r_2$ Å) | ($r_1$ Å) | $N_1$ | $n_1$ | $n_2$ | $n_3$ | $N_2$ | $n_1/N_1$ | $n_2/N_1$ | $n_3/N_1$ | $P_1$ | $P_2$ |
|---|---|---|---|---|---|---|---|---|---|---|---|
| **9.82126** | **8.3101** | **132** | **48** | **80** | **4** | **178** | **0.3636** | **0.6060** | **0.0303** | **50** | **89** |

**Radius 9Å Relaxed Stucture**

| ($r_2$ Å) | ($r_1$ Å) | $N_1$ | $n_1$ | $n_2$ | $n_3$ | $N_2$ | $n_1/N_1$ | $n_2/N_1$ | $n_3/N_1$ | $P_1$ | $P_2$ |
|---|---|---|---|---|---|---|---|---|---|---|---|
| 8.6727 | 7.4667 | 90 | 23 | 55 | 12 | 136 | 0.255 | 0.6111 | 0.133 | 38 | 69 |

**Radius 7Å Relaxed Stucture**

| ($r_2$ Å) | ($r_1$ Å) | $N_1$ | $n_1$ | $n_2$ | $n_3$ | $N_2$ | $n_1/N_1$ | $n_2/N_1$ | $n_3/N_1$ | $P_1$ | $P_2$ |
|---|---|---|---|---|---|---|---|---|---|---|---|
| 6.720 | 5.295 | 68 | 32 | 36 | 0 | 44 | 0.4705 | 0.5294 | 0 | 26 | 19 |

Where $r_2$ is maximum radial distance, $r_1$ is the radial distance below which bulk atmosphere remains. $N_1$ is the total number of Co atoms in ($r_2$-$r_1$), $N_2$ is the number of Co atoms with bulk atmosphere. $n_1$ is the number of Co atoms with 1 C NN, $n_2$ is the number of Co atoms with 2 C NN and $n_3$ is the number of Co atoms with 3 C NN in ($r_2$-$r_1$). $P_1$ is the total number of C atoms in ($r_2$-$r_1$) and $P_2$ is the number of C atoms with bulk atmosphere.